\documentclass[12pt]{article}
\usepackage{geometry}
\usepackage{amsmath,amssymb,setspace,fancyhdr,geometry,url}
\usepackage{graphicx,caption}
\usepackage{lscape,amsthm}
\usepackage{gensymb}
\usepackage{bm}
\usepackage{natbib}
\usepackage{titlesec}
\usepackage{natbib}
\usepackage{color} 

\usepackage{algorithm,algpseudocode}
\usepackage{caption}
\allowdisplaybreaks[1]
\allowbreak

\renewcommand{\thesection}{}
\renewcommand{\thesubsection}{}
\titleformat{\section}{\Large\bfseries}{\thesection}{0em}{}
\titleformat{\subsection}{\large}{\thesubsection}{1em}{}

\title{Quantifying Spatio-Temporal Variation of Invasion Spread}
\author{Joshua Goldstein$^1$, Jaewoo Park$^2$, Murali Haran$^{2*}$, Andrew Liebhold$^3$ and \\ Ottar N. Bj$\o$rnstad$^4$}
\date{}
\begin{document}

\doublespacing
\label{firstpage}

\maketitle

\noindent$^1$Social and Data Analytics Laboratory, 900 N Glebe Rd, Virginia Tech, Arlington, VA 22203 USA\\
$^2$Department of Statistics, Pennsylvania State University, University Park, PA 16802 USA\\
{*}email: mharan@stat.psu.edu\\
$^3$US Forest Service Northern Research Station, Morgantown, WV 26505 USA\\
$^4$Departments of Entomology and Biology, Pennsylvania State University, University Park, PA 16802 USA


\clearpage

\begin{abstract}\noindent
\begin{itemize}
\item The spread of invasive species can have far reaching environmental and ecological consequences. Understanding invasion spread patterns and the underlying process driving invasions are key to predicting and managing invasions.
\item We combine a set of statistical methods in a novel way to characterize local spread properties and demonstrate their application using simulated and historical data on invasive insects. Our method uses a Gaussian process fit to the surface of waiting times to invasion in order to characterize the vector field of spread.
\item Using this method we estimate with statistical uncertainties the speed and direction of spread at each location. Simulations from a stratified diffusion model verify the accuracy of our method.
\item We show how we may link local rates of spread to environmental covariates for two case studies: the spread of the gypsy moth ({\it Lymantria dispar}), and hemlock wolly adelgid ({\it Adelges tsugae}) in North America. We provide an R-package that automates the calculations for any spatially referenced waiting time data. 

\end{itemize}
\end{abstract}

\begin{center}{\bf Key-words:} invasive species, gypsy moth, hemlock wolly adelgid, Gaussian process, spatial gradients
\end{center}
\clearpage

\section{Introduction}\label{sec:intro}

~~~~When a non-native species successfully establishes in an exotic environment it enters the spread phase of biological invasions during which the species expands its range into suitable habitat \citep{lockwood2013invasion}. Ecological theory has shown that the speed of invasion spread is a joint function of the dispersal rate and the population growth rate of the invading species \citep{skellam1951random,okubo1980diffusion}; any habitat characteristic that influences population growth or dispersal can thus influence the rate of spread. Rates of spread may vary considerably among species and for a given species, spread rates may vary across heterogeneous landscapes \citep{shigesada1987speeds, tobin2007invasion}. Understanding the mechanisms causing heterogeneity in the rate of invasion spread is key to predicting future rates of spread and identifying important locations for management.

In this work we propose automated statistical methods for estimating local speed and dominant direction of spread along invasion fronts. Our approach can be applied to identify statistically significant environmental and geographic determinants of local invasion rates and likely epicentra of invasion resulting from long-range introductions.

In addition to environmentally-driven heterogeneity in rates of spread, there is considerable variation among species in the extent to which invasion spread is discontinuous("jumps"). Spread of some species occurs via continuous expansion of the range into contiguous areas. For example, the North American muskrat, {\it Ondatra zibethica}, invaded central Europe from 1905-1927 via gradual expansion of its range in concentric circles \citep{skellam1951random}.  The spread of other species is highly discontinuous, characterized by a pattern referred to as stratified diffusion \citep{shigesada1995}; following initial establishment, expansion may happen with long-range jumps into isolated uninvaded areas, founding new colonies that expand and eventually coalesce to form a contiguously invaded zone. This pattern is observed in many species of invading organisms, such as invasion of North America by the Argentine ant, {\it Linepithema humile} \citep{suarez2001patterns} and the gypsy moth, {\it Lymantria dispar} \citep{sharov2002slow}.

Quantifying the spread of non-native species and relating invasion speed to habitat heterogeneity is important for predicting and managing biological invasions. Several methods have been developed for studying processes that control spread rates of species. Species distribution models \citep{guisan2000predictive,stauffer2002linking,guisan2005predicting,elith2009species,elith2015predicting} are widely used to predict distributions of invasive species, for example by using generalized linear models or generalized additive models.  A variety of methods  \citep{wikle2003hierarchical,hooten2007hierarchical,hooten2008hierarchical,wikle2010general,hooten2010statistical,bled2011hierarchical,broms2016dynamic} combine dynamic equations within the framework of a hierarchical Bayesian model. These novel approaches embed dynamic equations within statistical models, allowing for a scientific interpretation of their fitted models. The above work has largely used spatial counts or presence-absence disease data; in contrast, the data we use is the time of first appearance of an invasive species. 

We note that there are numerous other ways to model data on the spread of invasive species, including data in the form of point-level spatial data \citep[cf.][]{latimer2009hierarchical, hanks2017modeling}.

Several methods have been developed for measuring spread based upon fitting range size to time since establishment or estimating spread by directly quantifying displacement of range boundaries over time \citep{sharov1997methods,tobin2007comparison,gilbert2010comparing}.  These methods are generally well-suited for quantifying average spread range and temporal variation therein, but they are limited in their ability to quantify local spread rates and their relation to local habitat characteristics.  Also, these methods are generally designed to quantify spread as a continuous process; identification of long-range jumps in stratified dispersal is usually done visually in a non-automated fashion. These gaps in existing methodology provides our motivation for combining recent developments in spatial statistics methodology in order to provide an automated approach to estimate local speed and direction of spread. Here our focus is on  constructing a spatial surface  that describes the direction and speed of spread  of an invasive species. Our method can help researchers learn about characteristics of the spread of the invasive  species, including both local speed and direction as well as long range. We take advantage of recent statistical theory on the estimation of spatial gradients. We test our methods on simulated data generated from a stratified diffusion model and apply them to two detailed case studies of biological invasions, the historical spread of the gypsy moth and the hemlock wolly adelgid, {\it Adelges tsugae}, in North America.

\section{Data} \label{sec:data}

\subsection{Gypsy moth}

~~~~Native to Europe and Asia, the gypsy moth was accidentally introduced from France to Massachusetts in the late 1860's \citep{liebhold1989learning}, it has since spread throughout much of the northeastern USA. The gypsy moth is now established in a large area composed of the north Atlantic states and bordering Canadian provinces, as well as a second focus resulting from a long-range jump event to Michigan around 1980 \citep{liebhold1992gypsy,johnson2006allee,tobin2007invasion}.

The invasion of the gypsy moth across North America has been slow compared to the rate of spread of many other alien species \citep{liebhold2008population}. Mean spread was estimated at 21 km per year from 1960 to 1990 \citep{liebhold1992gypsy}. The relatively slow rate of spread can be attributed, in part, to the fact that females of North America populations are flightless. Gypsy moth populations spread by short-range windborne dispersal of $1^{\text{st}}$ instar larvae through a process known as `ballooning' \citep{mason1981larval}. Egg masses are also accidentally transported across longer distances on wood or human-made objects, forming new colonies ahead of the invasion front and resulting in a pattern of stratified diffusion \citep{sharov2002slow}.

The full invasion history of the gypsy moth in the US is reflected in the year of government designation of gypsy moth quarantine by county. County-level quarantine records for gypsy moth are maintained by the United States Department of Agriculture (U.S. Code of Federal Regulations, Title 7, Chapter III, Section 301.45). Historically, an entire county was usually designated part of the quarantined area when established gypsy moth populations were first detected anywhere within the county. These records are updated annually and exist from 1934 to the present. From 1900 to 1934, the year when counties were first infested has been described in various other published sources \citep[e.g.,][]{ burgess1913report, burgess1915report, liebhold1992gypsy}.  As additional covariates, we used county-level data derived from a national forest inventory system on the percent of the forest basal area comprised of oaks, which is a favored food plant of the gypsy moth, and the size (km$^2$) of each county \citep{liebhold1997gypsy}.

\subsection{Hemlock wolly adelgid}

~~~~Hemlock woolly adelgid (HWA) is an insect species responsible for defoliation of its host trees, eastern hemlock and Carolina hemlock \citep{orwig2002landscape,morin2009anisotropic}. Native to East Asia, it was first discovered in the eastern USA in Virginia in the 1950s \citep{ward2004eastern}.  HWA life stages can be transported by wind, wildlife, especially birds, and humans. Since its discovery, it has gradually expanded its range into much of the northeastern USA \citep{evans2007geographically,morin2009anisotropic}. By 1969 it was found in southern Pennsylvania and it invaded southern New England by 1985, spreading at an estimated speed of 20-30 km/year \citep{morin2009anisotropic}.

As with the gypsy moth, historical spread of the HWA was recorded at the county level. Records from the US Forest Service Forest Health Protection are available for 1951, 1971, 1981, 1996, and from 2001 to 2011.   We use the basal area of hemlock \citep{morin2004mapping} and plant hardiness zone \citep{cathey1990usda} for each county as additional covariates for our analysis.

\section{Methods} \label{sec:methods}

~~~~Historical spread of the gypsy moth has previously been estimated as averages over space. \cite{liebhold1992gypsy} estimated spread rates for five geographic regions by the slope of a least-squares regression of time on distance to a reference point in each region. Spread rates have also been estimated by measuring the average displacement of range boundaries over time \citep{sharov1997methods,tobin2007comparison}.

Previous research on quantifying spatial gradients from georeferenced biological data has focused on detecting zones or boundaries of rapid change across space using geostatistical {\it wombling} \citep{womble1951}. Wombling methods involve estimating local vector gradients by fitting bilinear functions over a lattice of points. This method has been applied to genetic \citep{barbujani1989detecting} as well as ecological \citep{fortin1994edge} data. More recent wombling methods for areal data feature Bayesian hierarchical spatial models in order to identify significant boundaries after accounting for spatial dependence via Markov random fields \citep{banerjee2004hierarchical,fitzpatrick2010,lu2007bayesian}, with applications to ecology and epidemiology.

The use of spatial gradients to estimate biological spread is motivated by the fact that if the surface is the waiting time to first appearance, then the reciprocal of the gradient length is a measure of the invasion speed: Fast spread leads to shallow waiting time surfaces, while slow spread results in steep surfaces. Previously \cite{johnson2004landscape} estimated spread gradients using a thin plate spline applied to waiting times (as measured by wavelet phase angles) to study outbreak spatial dynamics of the larch budmoth. \cite{farnsworth2009} used a similar spline surface approach to study spread of avian influenza. The thin plate spline approach yielded gradients which reflect the magnitude and direction of the spread, a simple general-purpose approach for visualization, but does not yield measures of statistical uncertainty associated with local spread estimates which prevents rigorous inference regarding whether, for example, any observed spatial variation is significant. In order to facilitate understanding the models and inferential procedure, we summarize our approach in the  following sections. The mathematical details for the Gaussian process gradient models are provided in the supplement.

\subsection{Estimating gradient surface using Gaussian processes} \label{sec:models}


~~~~Given data on time of first appearance of an invasive species, we are interested in constructing a surface that describes the direction and speed of spread of the invasive species. We use Gaussian process models as a convenient and rigorous approach to estimate such a surface. Gaussian processes are commonly used for spatial interpolation \citep{kbiob1951statistical}. We use a Gaussian process to spatially interpolate time of first appearance. The gradient of this Gaussian process, which is known to also follow a Gaussian process \citep{banerjee2003directional}.


Based on fitting a Gaussian process to our data, we develop methods for estimating speed and direction of the spread of the invasive species, and for detecting sites of long-range dispersal. We also provide, in an electronic supplement, computer code for an R \citep{ihaka1996r, rmanual2013} software package that automates the inference.

\begin{figure}
\centering
\caption{Year of first appearance by county for the gypsy moth (a) and hemlock woolly adelgid (b).}
\includegraphics[width=\textwidth]{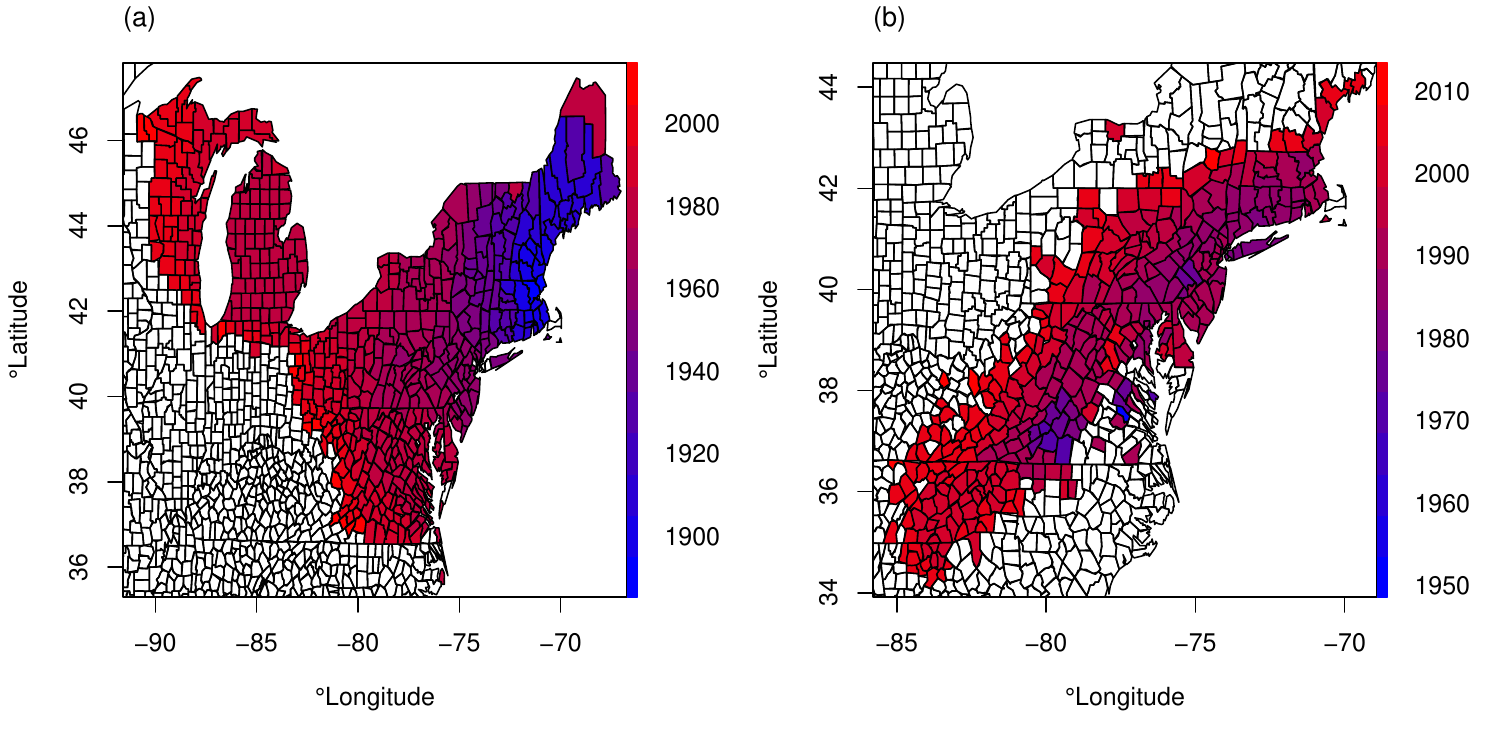}
\label{fig:first}
\end{figure}

We assume we have observations of the year of first appearance $\bm{Y} = \{ Y(\bm s_1),...,Y(\bm s_n) \}$ at locations $\{ \bm s_1,...,\bm s_n \}, \bm s_i \in \mathbb{R}^2$. For our examples, data are county-level quarantine records and the spatial locations $\{ \bm s_1,..,\bm s_n \}$ are taken to be the centroids of counties for the gypsy moth $(n=571)$ counties (Figure \ref{fig:first}a) and for the HWA $(n=340)$ counties (Figure \ref{fig:first}b). The data are discrete ("areal") in space as they represent counties. In order to use a Gaussian process gradient model, we treat the data as if they are from the centroid of each county. In order to investigate the potential sensitivity of our conclusions to this approximation, we perturb the locations of the centroids of each county and perform the analysis with this perturbed data. We find that the estimated spread patterns of the perturbed datasets are similar to those of the original dataset (see supplement for details). Coordinates are projected using the Albers equal area conic projection with standard parallels $29^{\degree}30'$ and $45^{\degree}30'$. $Y(\bm s_i)$ is the year county $i$ was added to the quarantine. We assume $Y(\bm s)$ can be modelled using an isotropic Gaussian process. For our applications, we assume the original process $Y(\bm s) = \mu(\bm s) + w(\bm s) + \epsilon(\bm s)$, with mean function  $\mu(\bm s) = \beta_0 + \beta_1 s_x + \beta_2 s_y$, correlated spatial error $w(\bm s) \sim GP(0, K(\cdot))$ with Mat\'ern covariance smoothness $\nu = 3/2$, which takes the explicit form $K(r)=\sigma^2 (1+\phi r) \text{exp}\{-\phi r\}$, and uncorrelated error $\epsilon(\bm s) \sim N(0,\tau^2)$, where $\tau^2$ is a nugget effect that captures measurement error.

The gradient of waiting time $\nabla Y(\bm{s})$ can be defined by taking the derivative of $Y(\bm{s})$ with respect to spatial directions over $\mathbb{R}^2$. The spatial gradient vector $\nabla Y(\bm{s}) \in \mathbb{R}^2$ indicates the dominant direction of spread. When $Y(\bm{s})$ is the time of first appearance of the species, the gradient length $\| \nabla Y(\bm{s}) \|$ measures the change in waiting time for spread of the species. Small change in time surfaces means fast spread, while large change indicates slow spread. Therefore, the reciprocal of the gradient length $1/\| \nabla Y(\bm{s}) \|$ represents the speed of spread. Because $Y(\bm{s})$ is a Gaussian process, well established results \citep{banerjee2003directional}  show how we can obtain the distribution of $\nabla Y(\bm{s})$ by using its direct relationship to the distribution of $Y(\bm{s})$. This allows us to estimate both the direction and speed of spread of the invasive species based on the observations of time of first appearance.

Our other interest is in detecting long range jumps. For each spatial location, our goal is to  investigate whether they represent plausible introduction well ahead of the general spatial diffusion. For this we utilize the concept of "total gradient" function, $\Gamma(r)$. For a particular location and for a given cardinal direction, the total gradient $\Gamma(r)$ measures the change in the waiting time for the spread of the species to a distance $r$ away from the current location. Small $\Gamma(r)$ means shallow time surfaces which comes from fast spread of the species. This implies a potential long range spread in that direction. Because $Y(\bm{s})$ is a Gaussian process, we can easily also obtain the distribution of $\Gamma(r)$ \citep{banerjee2003directional}. Based on this result, we can learn about the conditional distribution of $\Gamma(r)|Y(\bm{s})$ to search for any such long ranges jumps.

In addition to total gradient, we also investigate the use of a Rayleigh test from circular statistics \citep{jammalamadaka2001topics}. Although we find that this test is not a perfect method, it may still be a useful fast preliminary test for long range jumps. Details for the Rayleigh test are provided in the supplementary material.



\subsection{Inferential Procedure}\label{sec:longrange}

~~~~Our approach combines well established spatial statistics tools in a novel way. Our inferential procedure is based on the Gaussian process gradient model and may be summarized as follows.

\begin{enumerate}

\item The Gaussian process model is fit to $Y(\bm{s})$:\\
We infer the mean and covariance parameters $\Theta = ( \beta_0, \beta_1, \beta_2, \sigma^2, \phi, \tau^2 )$ of the Gaussian process $Y(\bm{s})$ based on a Bayesian approach. $\Theta$ is sampled from the posterior distribution using a Markov chain Monte Carlo (MCMC) algorithm. The posterior mean is estimated as $\widehat{\Theta} = \frac{1}{m}\sum_{i=1}^{m} \Theta_{i}$.

\item Detecting diffusive expansion:\\
We are interested in learning about local speed and direction of spread. For each location $\bm{s}_{i}$ and a given posterior sample $\Theta$, the gradient $\nabla Y(\bm{s}_{i})$ has the distribution  $\nabla Y(\bm{s}_{i})|Y(\bm{s}_{i}), \Theta$, which is a normal distribution because $Y(\bm{s}_{i})$ is modeled as a Gaussian process.

\begin{itemize}
\item The mean speed of spread is estimated as $\frac{1}{n}\sum_{i=1}^{n} 1/\| \nabla Y(s_i) \|$.
\item By plotting all statistically significant gradients (Figure \ref{fig:mothspread}) we can visualize the vector field of spread. 
\end{itemize}

\item Detecting sources and long-range jumps:\\
For each location $\bm{s_{i}}$ and a given posterior mean $\widehat{\Theta}$, we obtain the total gradient $\Gamma(r)$ from the conditional distribution $\Gamma(r)|Y(\bm{s_{i}}),\widehat{\Theta}$ which also follows a normal distribution.

\begin{itemize}
\item We flag a location as a potential site of a long-range introduction (Figure \ref{fig:mothspread}) if: (i)  the spread is significant  for at least two out of the four cardinal directions, and (ii) for the remaining directions it is not significantly small.  
\end{itemize}

\end{enumerate}

\subsection{Driving factors of spread}

~~~~We can gain insight into drivers of spread by relating the geographic variation in spread to habitat characteristics. To account for spatial dependence we fit a Bayesian spatial regression model to log-speeds using the {\it spBayes} $\texttt{R}$ package \citep{spBayes}. We apply a log transformation to the response since the speeds have right-skewed distributions. If the mean speed at location $\bm s_0$ is given by $V(\bm s_0)$, then we assume
\[
\text{log} V(\bm s_0) = X^T(\bm s_0) \bm\beta + w(\bm s_0) + \epsilon(\bm s_0)
\]
where $X( \bm s )$ is a vector of the spatially varying environmental and geographical covariates of interest. We assume $w(\bm s) \sim GP(0,G(\cdot))$, $G(\cdot)$ has Mat\'ern covariance smoothness with smoothness $\nu$, range $\phi$ and partial sill $\sigma^2$ and $\epsilon(\bm s) \sim N(0,\tau^2)$. Priors are selected as before and joint estimation is done via MCMC for  $\Theta = \{ \bm\beta, \sigma^2, \phi, \tau^2, \nu \}$.

\section{Results}\label{sec:results}

\subsection{Gypsy moth}

\begin{figure}
\centering
\caption{(a) Patterns of spread of the gypsy moth. Blue and red arrows indicate local speeds and directions of spread, and are plotted where spread is significant. The length of the arrows indicates the speed of spread -- longer arrows indicate faster spread. The color of the each arrow represents the time of first appearance of the process. Blue implies the earliest appearance, and red indicates the latest appearance. Green points indicate potential sites of long-range jumps. Green arrows around a point indicate significant directions of long range jumps. (b) Zoomed in figure of northeastern US.}
\includegraphics[width=\textwidth]{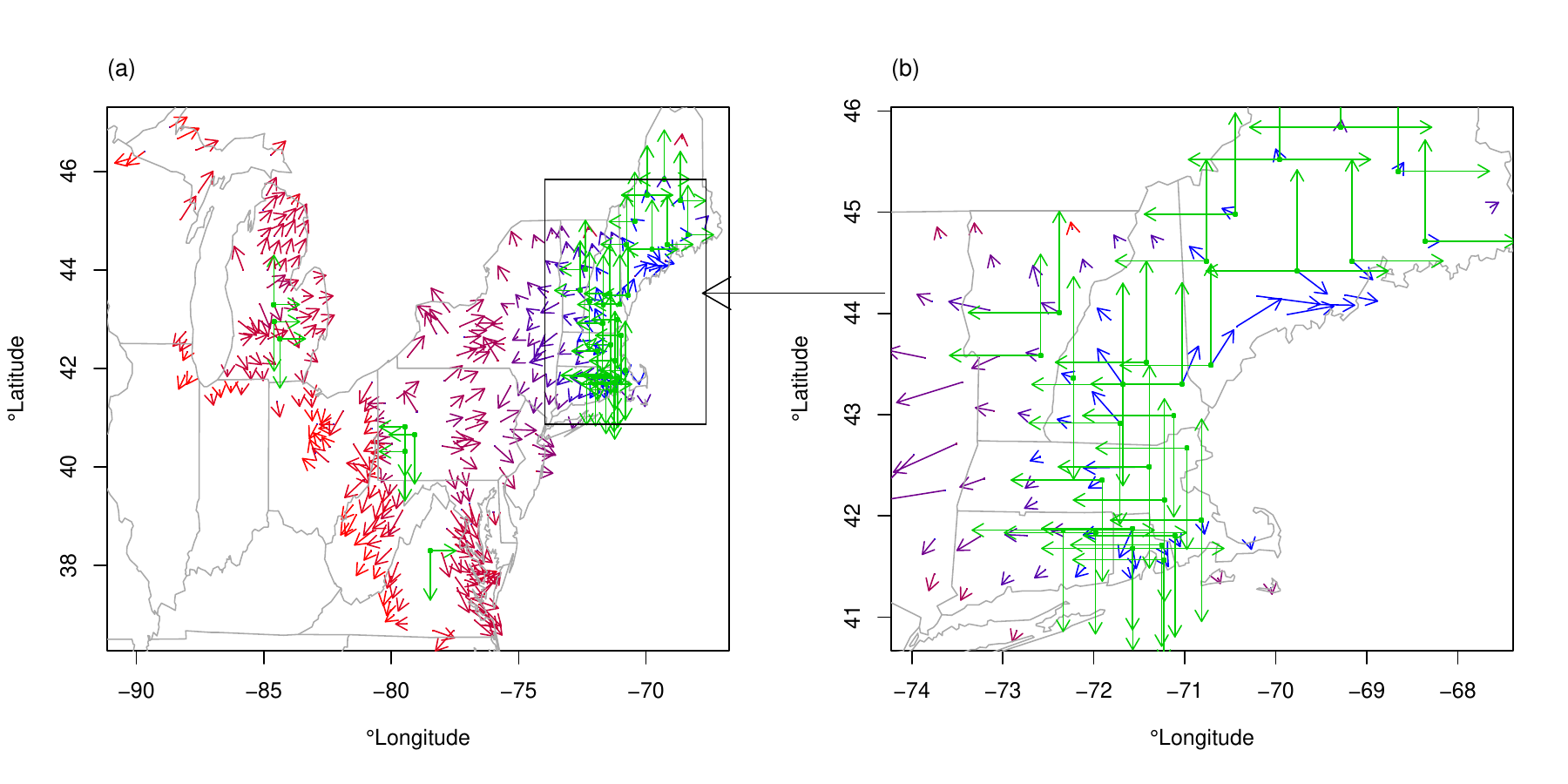}
\label{fig:mothspread}
\end{figure}

~~~~Significant speeds and directions of historical spread of the gypsy moth are plotted at the locations of each invaded county in Figure \ref{fig:mothspread}. The mean speed over all counties is 22.6 km/year, with a median of 15.7 km/year.  Distributions for the magnitude of spread at each location tend to be right skewed, where the 95\% credible interval is (1.7 km/year, 64.9 km/year).

In Figure \ref{fig:mothspread} we also test whether there are long range jumps of length $r=1^{\circ}$ in the four cardinal directions. Points identified as probable long range jumps are marked in green in Figure \ref{fig:mothspread}, along with green arrows which indicate significant directions of jumps. Our method identifies three potential sites around the northeastern coast, Michigan and central-western Pennsylvania. Prior analysis confirms two of these sites, as the population was first introduced in Massachusetts in the 1860s and a discrete population was later established in Michigan \citep{liebhold1992gypsy}. A close examination of Figure \ref{fig:first} also highlights a jump to Centre County, PA in the mid 1970s.

We relate speed of spread to latitude and longitude, quarantine date, county size, and finally the percent basal area comprised of trees preferred as hosts of the gypsy moth. Estimated parameters of the spatial regression model are given in Table \ref{table:regression}a. We verify that on average the gypsy moth spread faster as it moved west. We also found that basal area of susceptible host trees is significantly associated with faster invasion, consistent with the concept that local growth rates will be larger in the face of more favorable habitat, and should consequently enhance invasion spread rates.

\subsection{Hemlock woolly adelgid}

\begin{figure}
\centering
\caption{(a) Patterns of spread of the hemlock wolly adelgid. Blue and red arrows indicate local speeds and directions of spread, and are plotted where spread is significant. The length of the arrows indicates the speed of spread -- longer arrows indicate faster spread. The color of the each arrow represents the time of first appearance of the process. Blue implies the earliest appearance, and red indicates the latest appearance. Green points indicate potential sites of long-range jumps. Green arrows around a point indicate significant directions of long range jumps. (b) Zoomed in figure of Richmond area. }
\includegraphics[width=\textwidth]{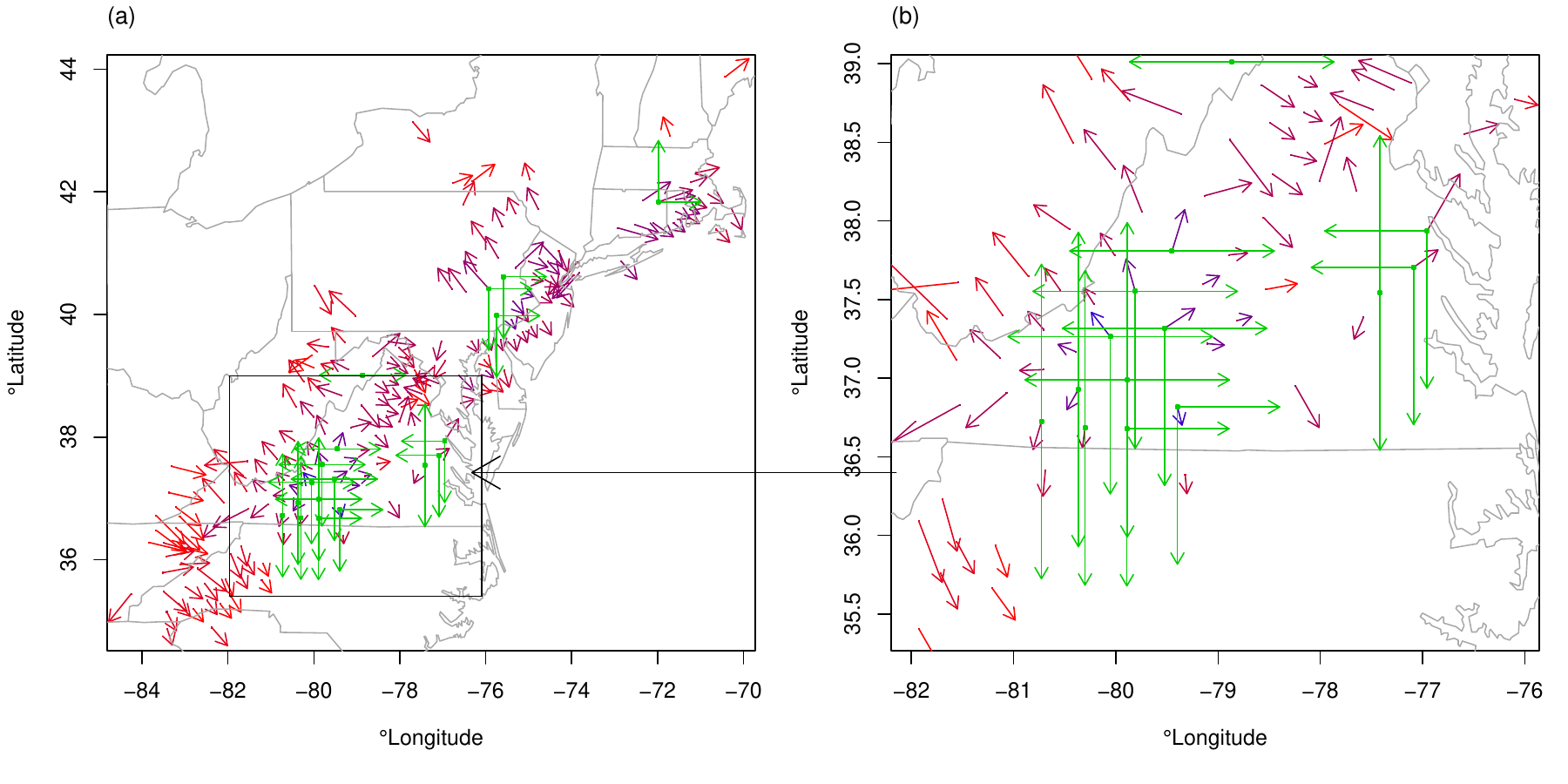}
\label{fig:hwaspread}
\end{figure}

~~~~Significant speeds and directions of spread for the HWA are plotted at each county in Figure \ref{fig:hwaspread}. We find a mean speed of spread of 20.5 km/year across counties, with a median speed of 13.5 km/year. Distributions for the magnitude of spread at each location tend to be right skewed, where the 95\% credible interval is (3.0 km/year, 59.2 km/year).

Probable sites of long-range introductions are also identified in Figure \ref{fig:hwaspread}. We detect areas of apparent long-range dispersal near Richmond, VA and southern PA, suggesting a pattern of stratified diffusion also for this species. \cite{morin2009anisotropic} previously found that expansion is significantly influenced by availability of host trees. Low winter temperatures can cause extensive mortality in HWA populations and limit expansion to the north \citep{trotter2009variation}. Therefore we relate speeds of spread to environmental features including the presence or absence of hemlock trees, and the average plant hardiness zone for each county, an index based on the mean annual minimum winter temperature \citep{cathey1990usda}. Estimates from the regression model are given in Table \ref{table:regression}b. We observed evidence that historically expansion is faster to the west and north. We also find as in \cite{morin2009anisotropic} that spread is significantly associated with the abundance of host trees. We also tested the interaction between plant hardiness zone and latitude and found that for a given latitude, HWA spread significantly slower through areas with lower (colder) plant hardiness zones $[ \beta = 3.4 (0.4, 6.3) ]$.

\subsection{Simulation}

~~~~We tested the ability of our method to recover the effects that spatially varying habitats have on the speed of spread. To accomplish this, data are simulated from a stratified diffusion model following \cite{shigesada1995}. Stratified diffusion is a combination of neighborhood diffusion and long distance dispersal. As the size of the original colony expands, new colonies are more likely to be created by long distance migrants.

The simulation starts with a single colony, centered at the initial point of invasion. The occupied area grows out a circle with the radius $r$ growing at constant rate $c$. This colony can then form offspring colonies from long-distance migrants in a random direction at a distance $L$ from the invasion front. New colonies form at a rate $\lambda(r)$ that is a function of the colony radius. These offspring colonies grow at speed $c$ and form offspring colonies of their own. The stratified diffusion simulation approach may be summarized as follows.

\begin{algorithm}
\caption{The stratified diffusion simulation approach}\label{stratified}
\begin{algorithmic}
\normalsize
\State Initialize with the first colony with the coordinates $\bm{s}_{0}$ and radius $r_{0}$.

\For{\texttt{$t=1:T$}}\\
~~~~ Given $n$th colony $\bm{s}_{n}$ with radius $r_{n,t}$ at time $t$.\\

~~~~ 1. Obtain the radius $r_{n,t+1}$ from $n$th colony: $r_{n,t+1} = r_{n,t} + cdt$, where $dt$ is\\ 
~~~~~ a time difference. (e.g. $dt = t+1 - t =1$)\\

~~~~ 2. With probability $\lambda(r_{n,t+1})$, a new colony $\bm{s}_{n+1}$ is generated in a random direction at a distance $L$ 

\EndFor\\

Return coordinates for $N$ number of simulated colonies $(\bm{s}_{0},\dots,\bm{s}_{N})$. Note that $N$ may be much smaller than $T$ if $\lambda$ is small. 
\end{algorithmic}
\end{algorithm}

We begin with an initial introduction in Massachusetts in 1900. Colony range expansion $c$ varies by longitude to simulate a slow period of initial expansion; $c=10$ km/year east of $-78^{\degree}$ and $c=20$ km/year west of $-78^{\degree}$. New colonies form at rate $\lambda(r) = 0.1 r$ a distance $L=10$ km from the invasion front. Additionally, to mimic the observed Gypsy Moth data an artificial long-range jump is introduced in Michigan in 1950. The simulation is run for $107$ years with an annual timestep.

\begin{figure}[h!]
\centering
\caption{(a) Patterns of spread of the simulated invasion. Blue and red arrows indicate local speeds and directions of spread, and are plotted where spread is significant. The length of the arrows indicates the speed of spread -- longer arrows indicate faster spread. The color of the each arrow represents the time of first appearance of the process. Blue implies the earliest appearance, and red indicates the latest appearance. Green points indicate potential sites of long-range jumps. Green arrows around a point indicate significant directions of long range jumps. (b) Waiting times of the stratified diffusion simulation \citep{shigesada1995}.}
\label{fig:simspread}
\includegraphics[width=\textwidth]{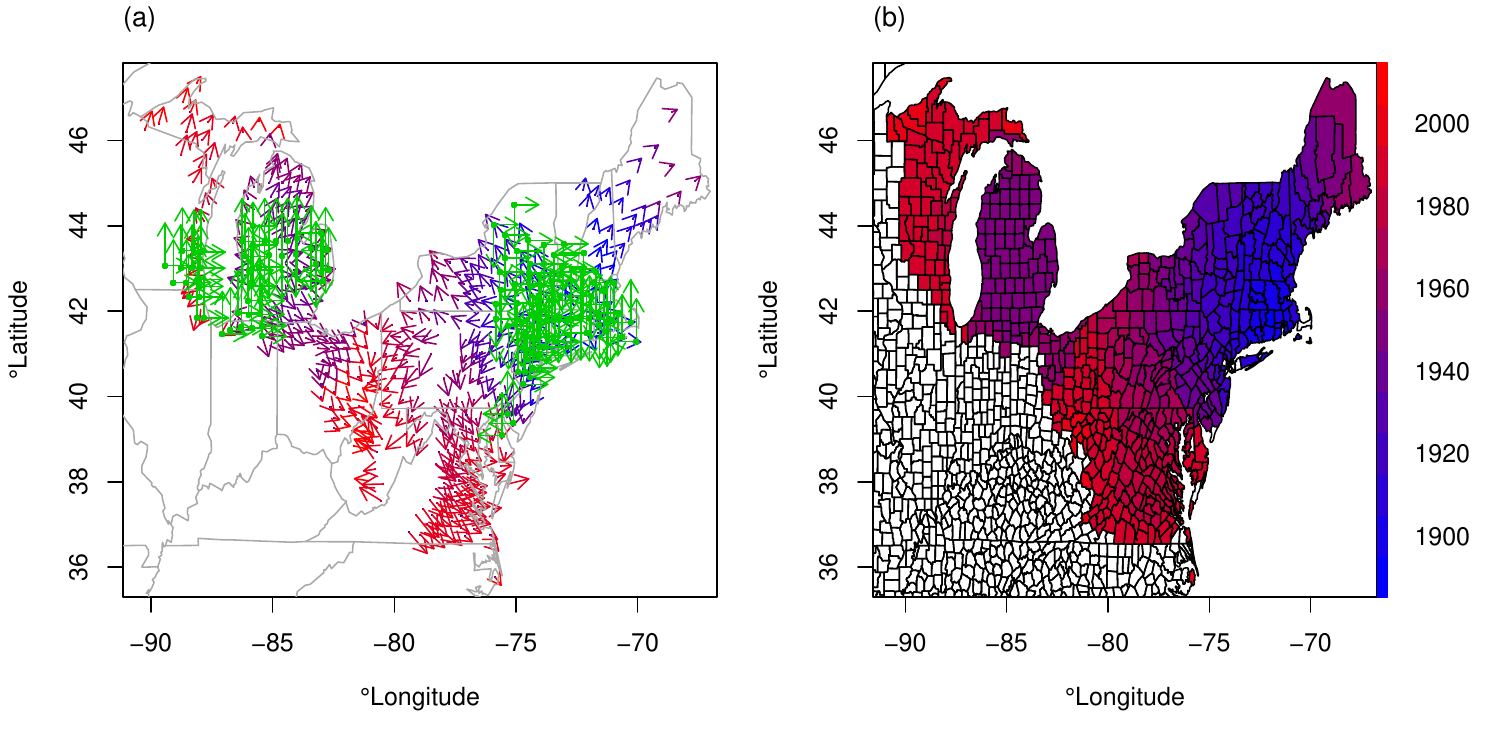}
\end{figure}

The time until the invasion front reaches each county is recorded as the simulated quarantine data (Figure \ref{fig:simspread}b). Figure \ref{fig:simspread}a indicates that our automated method successfully identified the two fixed colony introductions as regions of long-range jumps. We recover mean spread rates in the west of 10.7 km/year and in the east of 21.4 km/year, close to the true values used in the simulation. We also test our method under two different simulation scenarios -- slow spread and fast spread of the invasive species. Our method successfully detect long range jumps and recover the true spread rates well under both scenarios (see supplement for details).

\section{Discussion}\label{sec:discussion}

~~~~To study the establishment and spread of biological invasions we present a new method to estimate local rates and direction of spread, and identify key spatial features including sources, sites of rapid spread and long-range jumps. We visualize and make inferences on historical patterns of spread of the gypsy moth and hemlock woolly adelgid as well as validate the methodology on simulated data. Posterior inference in a Bayesian setting allows us to test the significance of spread patterns and spatial features of these invasions in a statistically rigorous way.

Taking our local estimates of gypsy moth spread and averaging them across time yields results in line with previous estimates \citep{liebhold1992gypsy}. We find an average speed of 11.4 km/year across counties quarantined from 1900 to 1915, followed by a slow spread (5.0 km/year) across counties from 1916 to 1965 and then a period of very rapid expansion (25.8 km/year) from 1966 to 2000. These changes may also be related to the differences in Allee effects among different regions along the invasion front as evidenced in \cite{tobin2009role}. From 2000 to present, coincident with USDAs ``Slow the Spread'' program of control \citep{sharov2002slow} we calculate an average speed of 14.6 km/year.

Our estimates for the spread of HWA when spatially averaged are also in line with previous estimates (e.g. \cite{ward2004eastern}). There is evidence that HWA range expansion is limited both by a lack of host trees and in the north by winter temperatures. 
 We note the important role of wildlife, especially migratory birds, as a means for HWA movement \citep{mcclure1990role}, in addition to windborne dispersal and human transport \citep{morin2009anisotropic}.

Our abilities to identify patterns of spread are constrained by the spatial and temporal resolution of our data. County-level quarantine data are typically coarser than, for example, gypsy moth pheromone trap count data, though \cite{tobin2007comparison} showed the two sources of gypsy moth data provided similar spread estimates. Additionally, the original Gaussian process must be sufficiently smooth for a gradient process to exist (we take the Matern model with smoothness parameter $\nu=3/2$), with the consequence that some information is lost at lcoal scales. We rely for the most part on annual records, but before 2001 the range of HWA was recorded at less frequent intervals. This is a potential source of bias in our early analysis of HWA spread.

For large spatial datasets, fitting a Gaussian process is a computational burden. Once the original Gaussian process is fit, however, we can draw samples by composition from the gradient process quickly. When the number of spatial locations is in the thousands we must likely have to rely on approximations such as the predictive process model of \cite{banerjee2008gaussian}.

Generally whenever the data are point-referenced waiting times, the speeds of spread can be estimated from the inferred gradient process. Therefore the methods presented here should be generally applicable to both ecological and epidemiological invasions. These methods are also potentially applicable to non-invasion problems such as the spread of an advantageous allele \citep{fisher1937wave}, or recurrent outbreak waves \citep{johnson2004landscape}. An \texttt{R} package that automates the inference is available as an electronic supplement.

\section{Acknowledgements}

This work has been funded by the Bill and Melinda Gates Foundation and by the National Science Foundation, \#DEB-1354819.

{\it{Competing interest}}: we have no competing interests.

\section{Data Accessibility}

We will put the data and code in the following repository.\\
(http://www.personal.psu.edu/muh10/invasionSpeed.html)

\clearpage

\begin{table}
\centering
\caption{Results of a spatial regression of speeds of spread (km/year) for the gypsy moth (a) and hemlock wolly adelgid (b) including posterior means and 95\% credible intervals obtained using the highest posterior density interval algorithm \citep{chen2000hpd}.}
\begin{tabular}{c|c}
{\bf (a) Gypsy Moth} & $\beta$ \\
\hline
Intercept & $-1.6 (-11.0,9.3)$ \\
Longitude & $-5.1 (-8.1,-2.2)$ \\
Latitude & $-2.3 (-6.6,1.2)$ \\
County size & $-0.00007 (-0.00020,0.00002)$ \\
Quarantine date & $0.0006 (-0.0044, 0.0056)$ \\
Basal\% susceptible trees & $0.0023 (0.0000, 0.0042)$ \\
\hline
{\bf (b) HWA} & $\beta$ \\
\hline
Intercept & $19.5 (3.1,36.6)$ \\
Longitude & $-9.8 (-14.9,-4.8)$ \\
Latitude & $8.5 (1.7,16.0)$ \\
Quarantine date & $-0.003 (-0.009,0.003)$ \\
I$_{\text{presence of hemlock}}$ & $0.09(0.01,0.07)$ \\
Plant hardiness zone & $0.014 (-0.19,0.23)$ \\
\end{tabular}
\label{table:regression}
\end{table}

\clearpage

\clearpage

\doublespacing

\appendix
\begin{center}
\title{\LARGE{\textbf{ Supplementary Material for Quantifying Spatio-Temporal Variation of Invasion Spread}}}\\
\author{Joshua Goldstein, Jaewoo Park, Murali Haran, Andrew Liebhold, and \\
Ottar N. Bj$\o$rnstad}
\end{center}

The supplementary material provides details about Gaussian process gradient models and simulation studies under different scenarios. 

\section{~Gaussian Process Gradient Models}

~~~~We assume we have observations of the year of first appearance $\bm{Y} = \{ Y(\bm s_1),...,Y(\bm s_n) \}$ at locations $\{ \bm s_1,...,\bm s_n \}, \bm s_i \in \mathbb{R}^2$. For our examples, data are county-level quarantine records and the spatial locations $\{ \bm s_1,..,\bm s_n \}$ are taken to be the centroids of counties for the gypsy moth where $n=571$ counties and for the HWA where $n=340$ counties. Coordinates are projected using the Albers equal area conic projection with standard parallels $29^{\degree}30'$ and $45^{\degree}30'$. For county $i$, $Y(\bm s_i)$ is the year the county was added to the quarantine. We assume $Y(\bm s)$ can be modelled using an isotropic Gaussian process with mean $\mu(\bm s)$ and covariance $K(\cdot)$.

\cite{banerjee2003directional} defines the finite difference directional derivative process at location $\bm s$ for scale {\it h} in direction $\bm u$ as
\[
Y_{\bm{u},h}(\bm{s}) = \dfrac{ Y(\bm{s} + h\bm{u}) - Y(\bm{s}) }{ h }
\]
where $\bm u$ is a unit vector. The directional derivative process in direction $\bm u$ is then defined as
\[
D_{\bm{u}} Y(\bm{s}) = \displaystyle\lim_{h \rightarrow 0} Y_{\bm{u},h}(\bm{s})
\]
assuming the mean square limit exists. Of interest is the gradient of $Y(\bm s)$ at $\bm{s}$, the vector of directional derivatives
\[
\nabla Y(\bm{s}) = \left( D_{\bm{e_1}} Y(\bm{s}), D_{\bm{e_2}} Y(\bm{s}) \right)
\]
in orthonormal basis directions $\bm{e_1} = (1,0)$ and $\bm{e_1} = (0,1)$. \cite{banerjee2003directional} shows there exists a joint trivariate Gaussian process for $Y(\bm{s})$ and $\nabla Y(\bm{s})$, and therefore $\bm Y$ and $\nabla \bm Y = \{ \nabla Y(\bm s_1),...,\nabla Y(\bm s_n) \}$ have a joint multivariate normal distribution which takes the form
\[
\left( \begin{array}{c}
\textbf{Y} \\
\nabla \textbf{Y} \\
\end{array} \right) \sim
N_{3n} \left[
\left( \begin{array}{c}
\boldsymbol{\mu} \\
\nabla \boldsymbol{\mu} \\
\end{array} \right),
\left( \begin{array}{cc}
K(D) & -\nabla K(D) \\
\nabla K(D)^T & -H_K(D) \\
\end{array} \right)
\right]
\]
where $D$ is the $n \times n$ matrix of pairwise Euclidean distances of \textbf{s}, and $K(D)$ represents the $n \times n$ matrix of $K(\cdot)$ applied element-wise to $D$. $\nabla \boldsymbol{\mu}$ is the length $2n$ vector
\[
\bigg( \dfrac{ \partial{\mu} }{ \partial{x} }(s_1),...\dfrac{ \partial{\mu} }{ \partial{x} }(s_n),\dfrac{ \partial{\mu} }{ \partial{y} }(s_1),...,\dfrac{ \partial{\mu} }{ \partial{y} }(s_n) \bigg)^T,
\]
$\nabla K(D)$ is the $n$ x $2n$ matrix
\[
\left( \begin{array}{cc}
\dfrac{\partial{K}}{\partial{x}}(D) &  \dfrac{\partial{K}}{\partial{y}}(D) \\
\end{array} \right),
\]
and $H_K(D)$ is the $2n$ x $2n$ matrix
\[
\left( \begin{array}{cc}
\dfrac{\partial^2{K}}{\partial{x^2}}(D) &  \dfrac{\partial^2{K}}{\partial{x}\partial{y}}(D) \\
\dfrac{\partial^2{K}}{\partial{x}\partial{y}}(D) & \dfrac{\partial^2{K}}{\partial{y^2}}(D) \\
\end{array} \right).
\]
The conditional distribution of the gradient is therefore given by
\[
\nabla \textbf{Y} | \textbf{Y},\Theta \sim
N_{2n} \left(
\nabla \boldsymbol{\mu} - \nabla K(D)^T [K(D)]^{-1} (\textbf{Y} - \boldsymbol{\mu}), -H_K(D) - \nabla K(D)^T [K(D)]^{-1} \nabla K(D) \right).
\]
where $\Theta$ is a vector of mean and covariance parameters for the Gaussian process. Letting $\delta = (s_0-s_1,...,s_0-s_n)$, the estimated gradient at a point $s_0$ is given by
\begin{equation}
\begin{split}
\nabla Y(s_0) | \textbf{Y},\Theta \sim
N_2 \big(
\nabla \mu(s_0) - \nabla K(\delta)^T [K(D)]^{-1} (\textbf{Y} - \boldsymbol{\mu}), \\
-H_K(0) - \nabla K(\delta)^T [K(D)]^{-1} \nabla K(\delta) \big).
\label{eq:gradpred}
\end{split}
\end{equation}
Note that for the gradient process to be well-defined, the original process must be mean square differentiable and all second order partial derivatives of $K$ must exist. This is the case, for instance, when $K$ is chosen to belong to the Matern family with smoothness parameter $\nu > 1$ \citep{stein1999interpolation}.

For our applications, we assume the original process $Y(\bm s) = \mu(\bm s) + w(\bm s) + \epsilon(\bm s)$, with mean function  $\mu(\bm s) = \beta_0 + \beta_1 s_x + \beta_2 s_y$, correlated spatial error $w(\bm s) \sim GP(0, K(\cdot))$ with Mat\'ern covariance smoothness $\nu = 3/2$, which takes the explicit form $K(r)=\sigma^2 (1+\phi r) \text{exp}\{-\phi r\}$, and uncorrelated error $\epsilon(\bm s) \sim N(0,\tau^2)$, where $\tau^2$ is a nugget effect capturing both mesurement error and microscale variability.

We infer the mean and covariance parameters $\Theta = ( \beta_0, \beta_1, \beta_2, \sigma^2, \phi, \tau^2 )$ based on a Bayesian approach using a Markov chain Monte Carlo (MCMC) algorithm. Flat prior distributions are assumed for mean parameters and inverse gamma priors are assumed for the partial sill $(\sigma^2)$ and nugget $(\tau^2)$ with shape parameter $2$ and scale parameter set to an approximate value from the empirical semivariogram. For the range $(\phi)$ a uniform prior is chosen with a support that allows the process to vary from low to high dependency. Simulation from the predictive distribution of the gradient can then be done by composition; given each posterior sample of $\Theta$, a sample for $\nabla Y(s_0)$ can be drawn from (\ref{eq:gradpred}).

At a given location $s_0$, posterior samples are the gradient at $s_0$ in the $x$ and $y$ directions. When the data are times of first appearance, steeper gradients correspond to slower speeds. Therefore posterior estimates of the gradient (including Bayesian credible intervals) allow us to make inferences on the local speed and dominant direction of spread. Therefore the speed of spread is the inverse of the magnitude of the posterior gradient $\| \nabla Y(s_0) \|$, and the dominant direction of the spread is in the direction of the gradient $\nabla Y(s_0) / \| \nabla Y(s_0) \|$.

\section{~Detecting sources and long-range jumps}

~~~~We test directly the distribution of the gradient around a point, which will allow us to check if there is significant radial expansion around that point. Define a curve $\mathcal{C}_{t^*} = \{ s(t): t \in [0,t^*] \}$, where $s(t) = \left( s_1(t), s_2(t) \right) \in \mathcal{R}^2$ and $s'(t)$ is the componentwise derivative. Let $\eta(s(t))$ be the unit vector normal to the curve at the point $s(t)$. The total gradient normal to $\mathcal{C}_{t^*}$ is
\[
\Gamma(t^*) = \int_0^{t^*} \langle \nabla Y(s(t)), \eta(s(t)) \rangle dv,
\]
where $v$ is the arc-length of the curve, $v(t^*) = \int_0^{t^*} \| s'(t) \| dt$, and so
\[
\Gamma(t^*) = \int_0^{t^*} \langle \nabla Y(s(t)), \eta(s(t)) \rangle \| s'(t) \| dt,
\]
In \cite{banerjee2006bayesian} it is shown that the distribution for the total gradient over a curve is a Gaussian process on $[0,T]$, $\Gamma(t^*) \sim GP( \mu_\Gamma(t^*), K_\Gamma(\cdot, \cdot) )$ with
\[
\mu_\Gamma(t^*) = \int_0^{t^*} \langle \mu(s(t)), \eta(s(t)) \rangle \| s'(t) \| dt
\]
\[
K_\Gamma(t_1^*, t_2^*) = \int_0^{t_1^*} \int_0^{t_2^*} \eta^T(s(t_1)) H_K\left[ s(t_2) - s(t_1) \right] \eta(s(t_2)) \| s'(t_1) \| \| s'(t_2) \| dt_1 dt_2
\]
where $\mu(\cdot)$ is the mean of the original process $Y(s)$ and $H_K(\cdot,\cdot)$ is the hessian of the covariance of $Y(s)$. \\

The conditional distribution of interest is
\[
\Gamma(t^*) | \textbf{Y},\Theta \sim N\big( \mu_\Gamma - \gamma^T_\Gamma(t^*) [K(D)]^{-1} (\textbf{Y} - \boldsymbol{\mu}),
K_\Gamma(t^*, t^*) - \gamma^T_\Gamma(t^*) [K(D)]^{-1} \gamma_\Gamma(t^*) \big)
\]
where for $j = 1,...,n$
\[
\gamma^T_\Gamma(t^*)_j = \text{cov}( \Gamma(t^*), Y(s_j) ) = \int_0^{t^*} \langle K \left[ s(t) - s(j) \right], \eta(s(t)) \rangle \| s'(t) \| dt.
\]
The terms $\gamma_\Gamma(t^*)$ and $K_\Gamma(t^*, t^*)$ are not available analytically, and must be computed using numerical integration. The {\it average} gradient normal to the curve $\mathcal{C}_{t^*}$ is simply the total gradient $\Gamma(t^*)$ divided by the arc-length.

Using this method centered on each location we can test the gradient normal to four sides of a box with sides of length $r$. As an heuristic we say if the spread is significantly {\it out} of at least two sides of the box, and does not go significantly {\it into} any side of the box, we will flag the location as a potential site of a long-range jump. We do this for a grid over the region of interest. In our figures, red lines indicate sides where there is a significant outward spread.

We also test radial spread around the point by using methods from circular statistics. The Rayleigh test \citep[see e.g.][]{jammalamadaka2001topics} is a statistical test of whether a circular distribution is random or non-random. When applied to the vectors of spread near a point, a non-random distribution implies a unified directional spread through that point. We take directions of spread in a neighborhood around each centroid and test if these directions are drawn from a uniform circular distribution. Say we have $n$ estimated directional vectors of spread $x_i, y_i$ in a neighborhood around a point. Let $r$ be the length of the mean vector from this sample, $r = \sqrt{\bar{x}^2 + \bar{y}^2}$. The test statistic is given by $R = 2 nr^2$ for the test of \\
\indent $H_0:$ Directional vectors are distributed randomly \\
\indent $H_a:$ Directional vectors are distributed non-randomly. \\
Under the null hypothesis $R$ will be $\chi^2$ distributed with $2$ degrees of freedom. If the test fails to reject the null this may contribute evidence that the directional distribution is uniform as would be the case for radial spread from the point. While this may occur because the location is the point source of a long-range jump, the Rayleigh test will also flag areas with vectors that are converging {\it to} the point (akin to an ecological sink) or are truly random. Therefore while it is useful for flagging potential sites of long-range jumps, it is not a perfect method because we need to visually distinguish between the three different scenarios that all lead to failure to reject the null. Therefore, we used the Rayleigh test as an additional methods to check long range jumps. One might be concerned about multiple testing issues. However our testing approach is already conservative since our null hypothesis is that there {\it is} a long range jump. That is, if we applied multiple test correction (e.g. Bonferroni correction), we get more long range jumps than we would otherwise.

\section{~Simulation Studies}

~~~~To validate our methods, we provide two different simulation studies. First, we simulate datasets from a stratified diffusion model \citep{shigesada1995} under different speed of spread scenarios to check whether our methods can recover the true parameter value used in the simulation. Second, we study the robustness of our methods by perturbing the locations of the centroids of the counties for the gypsy moth data. 
 
\subsection{Two Different Speed of Spread Scenarios}

~~~~We provide a summary of the results of simulation studies under two different scenarios. Data for both studies are simulated from a stratified diffusion model \citep{shigesada1995}. The simulation starts in Massachusetts in the year 1900. To mimic the observed gypsy moth data we introduce an artificial long-range jump in Michigan in 1950. We varied constant spread rate $c$ under two scenarios -- slow spread and fast spread. The rest of the simulation settings are identical to those in the manuscript. 

\begin{figure}[h!]
\centering
\caption{(a) Patterns of spread of the simulated invasion. Blue and red arrows indicate local speeds and directions of spread, and are plotted where spread is significant. The length of the arrows indicates the speed of spread -- longer arrows indicate faster spread. The color of the each arrow represents the time of first appearance of the process. Blue implies the earliest appearance, and red indicates the latest appearance. Green points indicate potential sites of long-range jumps. Green arrows around a point indicate significant directions of long range jumps. (b) Waiting times of the stratified diffusion simulation \citep{shigesada1995} for slow spread.}

\label{fig:simspread5}
\includegraphics[width=\textwidth]{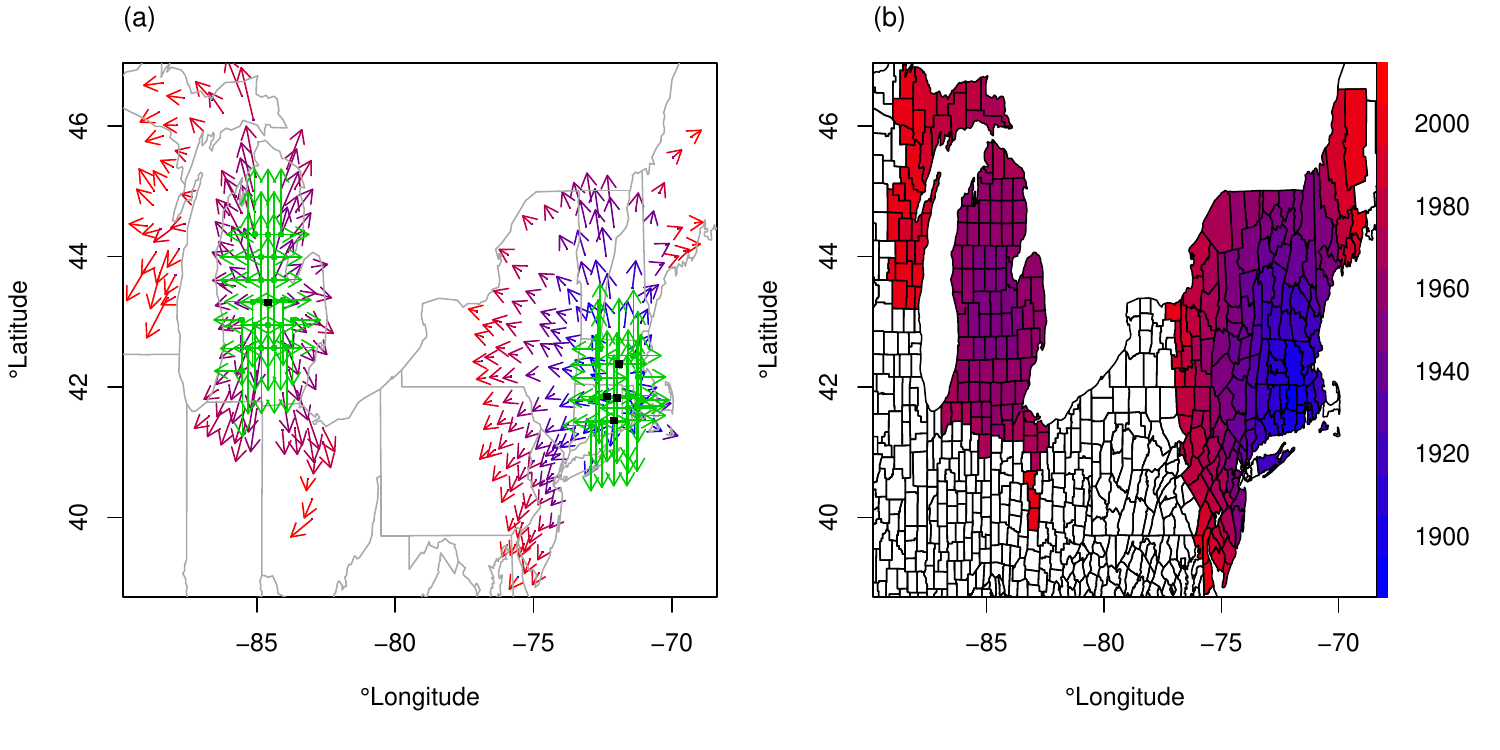}
\end{figure}

Compared to the simulation in the manuscript, we simulate a slow spread: $c=5$ km/year east of $-78^{\degree}$ and $c=10$ km/year west of $-78^{\degree}$. The time until the invasion front reaches each county is recorded as the simulated quarantine data (Figure \ref{fig:simspread5}b). In Figure \ref{fig:simspread5}a we observe that our method of inference has successfully identified the two fixed colony introductions as regions of long-range jumps. We recover mean spread rates of 4.9 km/year in the west and 10.1 km/year in the east. These are close to the actual values used in the simulation.

\begin{figure}[h!]
\centering
\caption{(a) Patterns of spread of the simulated invasion. Blue and red arrows indicate local speeds and directions of spread, and are plotted where spread is significant. The length of the arrows indicates the speed of spread -- longer arrows indicate faster spread. The color of the each arrow represents the time of first appearance of the process. Blue implies the earliest appearance, and red indicates the latest appearance. Green points indicate potential sites of long-range jumps. Green arrows around a point indicate significant directions of long range jumps. (b) Waiting times of the stratified diffusion simulation \citep{shigesada1995} for fast spread.}
\label{fig:simspread15}
\includegraphics[width=\textwidth]{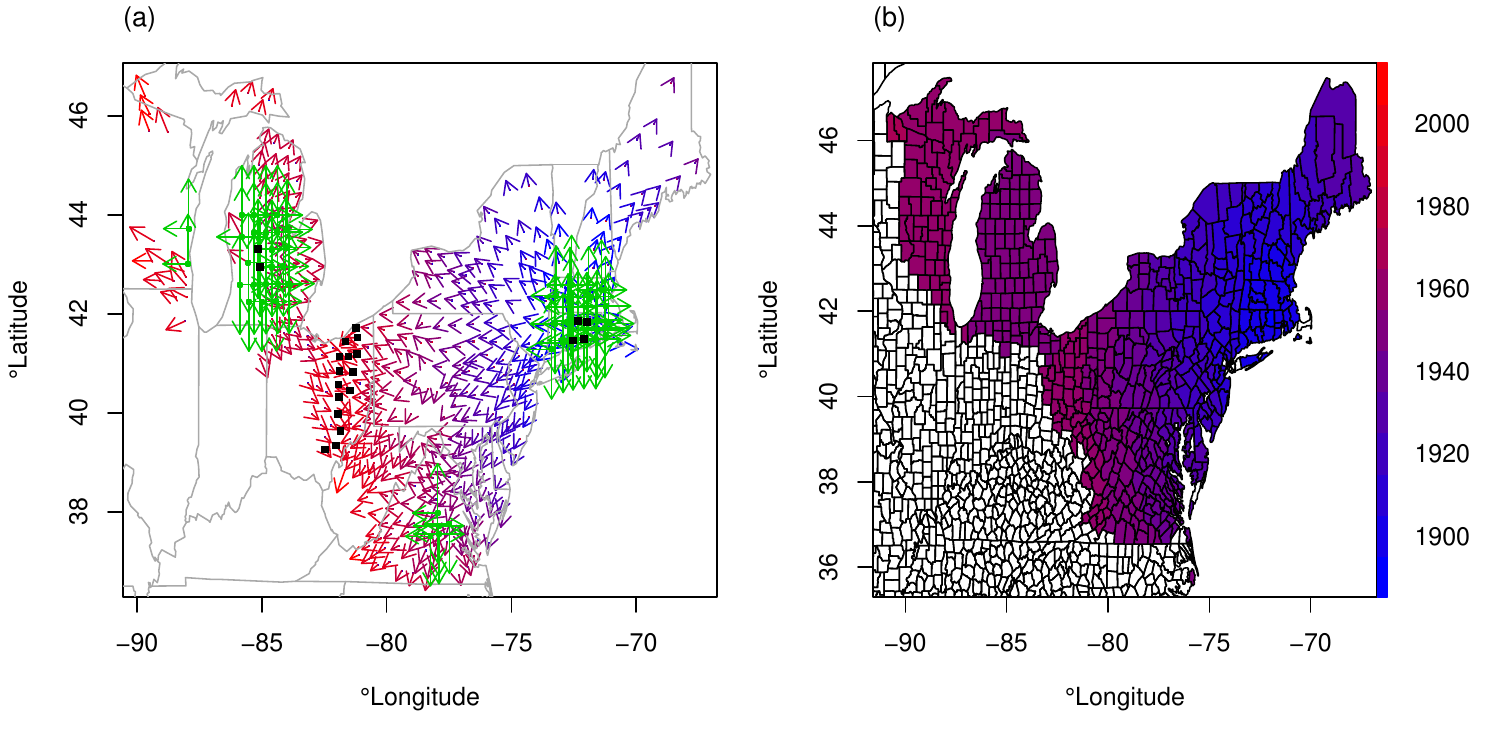}
\end{figure}

We also simulate a fast spread: $c=15$ km/year east of $-78^{\degree}$ and $c=30$ km/year west of $-78^{\degree}$. The time until the invasion front reaches each county is recorded as the simulated quarantine data (Figure \ref{fig:simspread15}b). In Figure \ref{fig:simspread15}a we observe that our method of inference has successfully detected the two fixed colony introductions as regions of long-range jumps. The mean spread rates recovered in the west and east are, respectively, 17.8km/year and  31.8km/year. These  are reasonably close to the actual values used in the simulation. 

\subsection{Perturbations of the Centroids of the Counties}

~~~~We are using continuous-space methods for data that are discrete (areal) in space. In particular, we treat the information that is county level as if it were information that was obtained at the centroid of each county. This is, of course, an approximation. To study the robustness of our approximate approach, we perturb the locations of the centroids of the counties for the gypsy moth data. For each coordinate, we add a Gaussian error with mean 0 and standard deviation proportional to the square root of the size of each county (i.e. $\bm s_i + N(0,0.01*\sqrt{size_i})$ ). In this way we generate three sets of perturbed data.

\begin{figure}[h!]
\centering
\caption{(a) Black dots represent the centroids of the counties in the gypsy moth data. Red dots indicate perturbed centroids. (b) Patterns of spread of the gypsy moth after perturbing the locations of the centroids of the counties. (c) Zoomed figure of northeastern US.}
\label{fig:perturb1}
\includegraphics[scale=0.39]{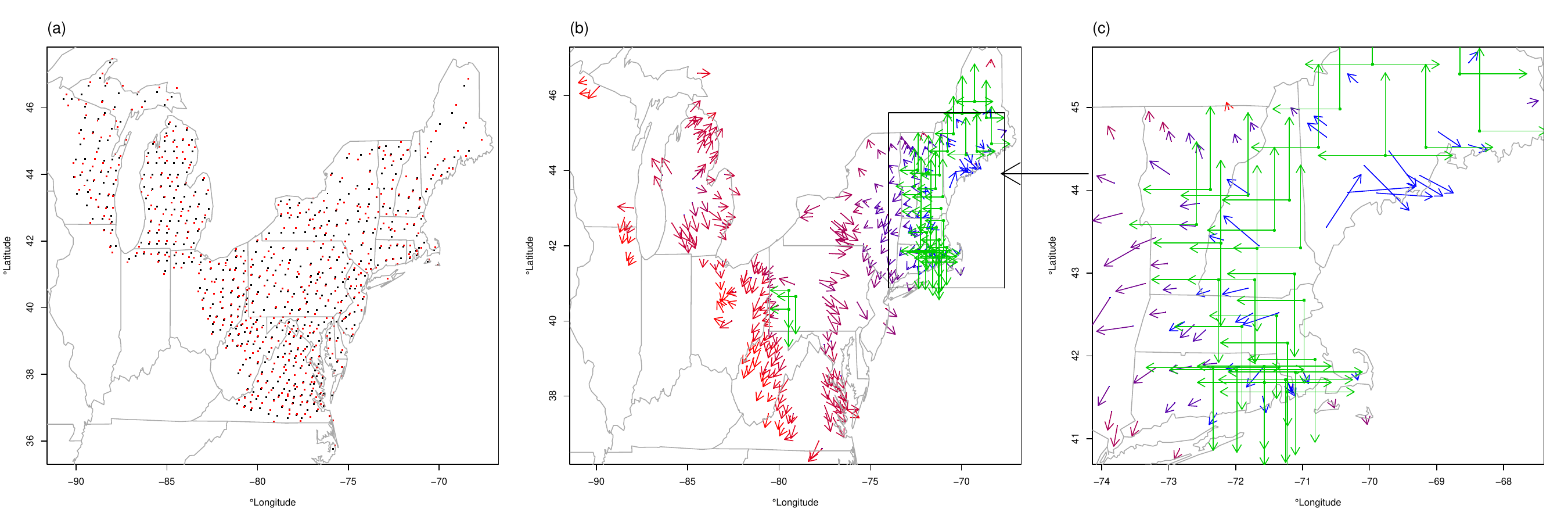}
\end{figure}

\begin{figure}[h!]
\centering
\caption{(a) Black dots represent the centroids of the counties in the gypsy moth data. Green dots indicate perturbed centroids. (b) Patterns of spread of the gypsy moth after perturbing the locations of the centroids of the counties. (c) Zoomed figure of northeastern US.}
\label{fig:perturb2}
\includegraphics[scale=0.39]{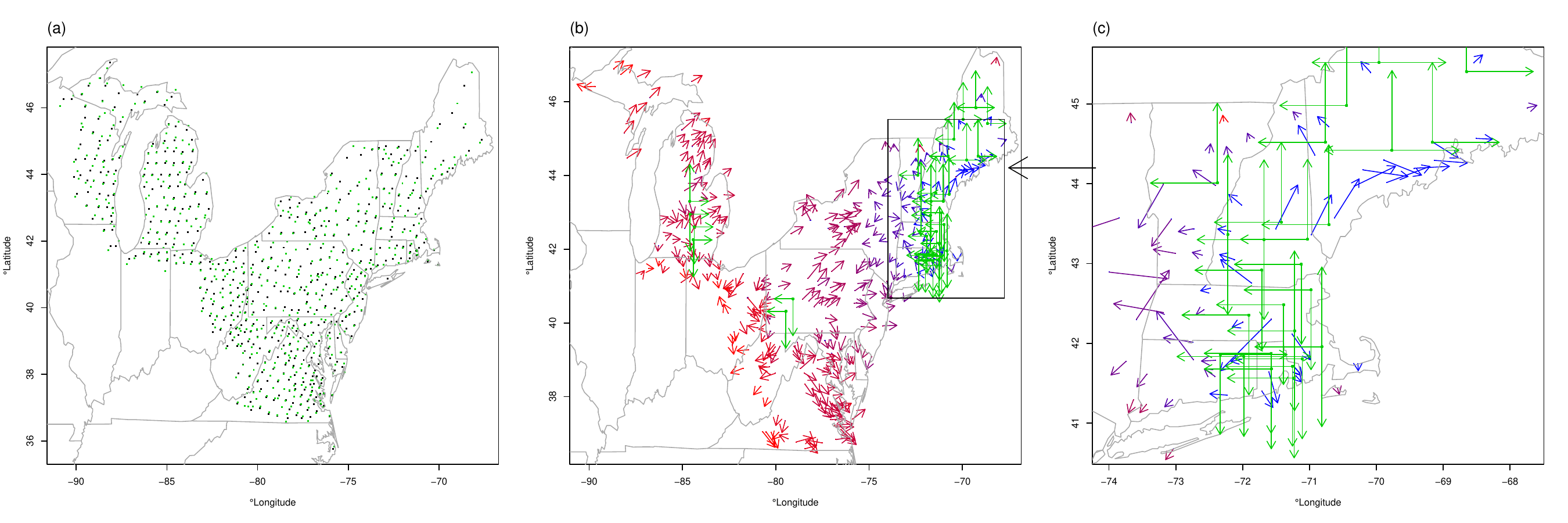}
\end{figure}

\begin{figure}[h!]
\centering
\caption{(a) Black dots represent the centroids of the counties in the gypsy moth data. Blue dots indicate perturbed centroids. (b) Patterns of spread of the gypsy moth after perturbing the locations of the centroids of the counties. (c) Zoomed figure of northeastern US.}
\label{fig:perturb3}
\includegraphics[scale=0.39]{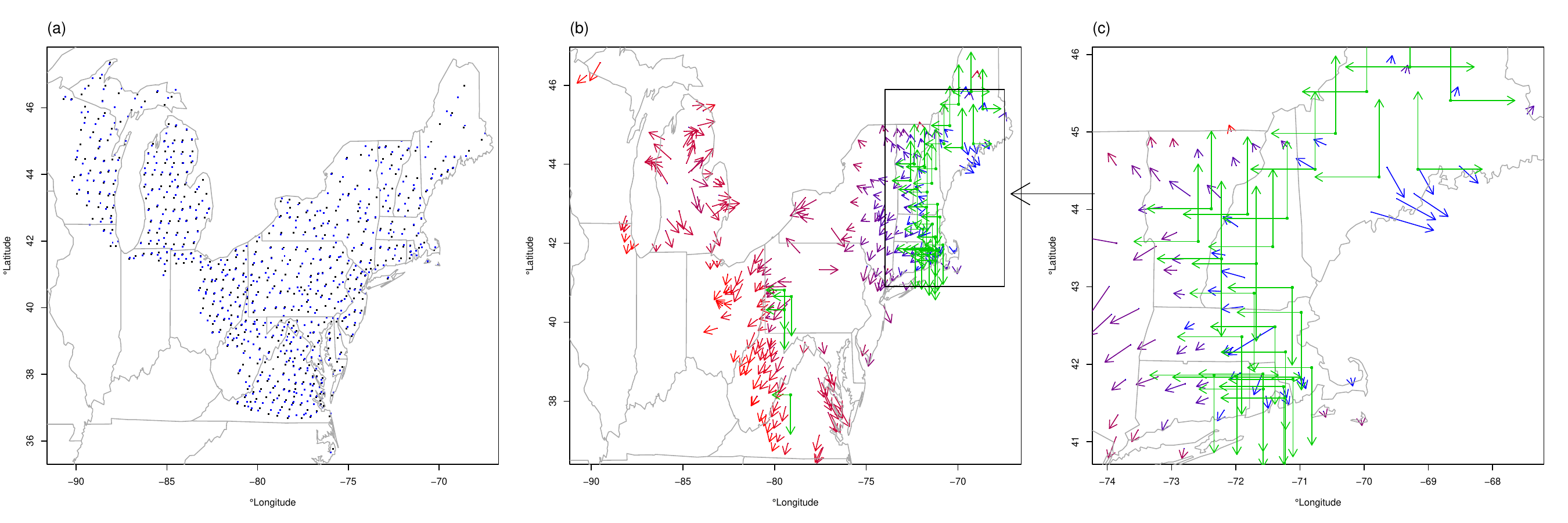}
\end{figure}

We then fit the Gaussian process gradient model to each perturbed data set. We observe that while there are differences in the inferred directions of significant spread, the overall patterns of spread for the perturbed data (Figure \ref{fig:perturb1}-\ref{fig:perturb3}) are similar to those of the original data (Figure 2 in the main paper). Furthermore, the mean speed of spread over all counties for three different perturbed data sets are estimated to be 22.3 km/year, 19.6 km/year, and 20.6 km/year, which are similar to the estimates for the original gypsy moth data, which is 22.6 km/year.

\clearpage

\clearpage

\clearpage

\clearpage

\clearpage

\bibliographystyle{apalike}
\bibliography{references}

\end{document}